\documentclass[useAMS,usenatbib]{mn2e}
\usepackage[dvips]{graphicx}
\usepackage{txfonts}
\usepackage{amssymb}
\usepackage{indentfirst}
\usepackage{epsfig}

\newcommand{\CVI}{\mbox{{C\,{\sevensize VI}}}}

\newcommand{\OVII}{\mbox{{O\,{\sevensize VII}}}}
\newcommand{\OVIII}{\mbox{{O\,{\sevensize VIII}}}}

\newcommand{\NeIX}{\mbox{{Ne\,{\sevensize IX}}}}

\newcommand{\Os}{\mbox{{O$^{5+}$}}}
\newcommand{\Ov}{\mbox{{O$^{6+}$}}}
\newcommand{\Oe}{\mbox{{O$^{7+}$}}}
\newcommand{\On}{\mbox{{O$^{8+}$}}}

\newcommand{\MgXI}{\mbox{{Mg\,{\sevensize XI}}}}

\newcommand{\FeXVII}{\mbox{{Fe\,{\sevensize XVII}}}}

\newcommand\foiii{f_{\rm OVIII}}
\newcommand\foii{f_{\rm OVII}}
\newcommand{\Lya}{\ensuremath{\hbox{Ly}\alpha~}}
\newcommand{\A}{\AA~}

\def\xmm{{\it XMM-Newton}}
\def\cha{{\it Chandra}}

\begin{document}
\title[Charge Exchange X-ray Emission of Star-forming Galaxies]{
		Charge Exchange X-ray Emission of Nearby Star-forming Galaxies}

\author[J. Liu et al.]{Jiren Liu$^{1,}$\thanks{E-mail: jirenliu@nao.cas.cn},
 Q. Daniel Wang$^{2}$ and Shude Mao$^{1,3}$ \\
$^{1}$National Astronomical Observatories,  Chinese Academy of Sciences, 20A Datun Road, Beijing 100012, China\\
$^{2}$Department of Astronomy, University of Massachusetts, Amherst, MA 01002, USA\\
$^{3}$Jodrell Bank Centre for Astrophysics, University of Manchester,
	  Manchester, M13 9PL, UK \\
}
\date{}

\maketitle

\begin{abstract}

Properties of hot gas outflows from galaxies are generally measured from associated
X-ray line emission assuming that it represents atomic transitions in thermally
excited hot gas. X-ray line emission, however, can also arise 
from the charge exchange between highly ionised ions and neutral species. 
The K$\alpha$ triplet of He-like ions can be used as a powerful diagnostic, because the 
charge exchange X-ray emission (CXE)
favours the inter-combination and forbidden lines, while the thermal emission
favours the resonance line. We analyse the \OVII\ triplet 
of a sample of nine nearby star-forming galaxies observed by the \xmm\
reflection grating spectrometers. 
For most galaxies, the forbidden lines are comparable to or stronger
than the resonance lines, which is in contrast to the thermal prediction.
For NGC 253, M51, M83, M61, NGC 4631, and the Antennae galaxy, the observed line 
ratios are consistent with the ratio of the CXE; for 
M94 and NGC 2903, the observed ratios indicate multiple origins; for M82,
different regions show different line ratios, also indicating multiple origins.
We discuss other possible mechanisms that can produce a relatively
strong forbidden line, such as a collisional non-equilibrium-ionization recombining/ionizing
plasma, which are not favoured. 
These results suggest that the CXE may be a common process 
and contribute a significant fraction of the soft X-ray line emission
for galaxies with massive star formation. 

\end{abstract}


\section{Introduction}

Galaxies with massive star formation are the primary source that 
ejects metals into galactic halos and the intergalactic medium. The properties 
of outflows driven by SN explosions and stellar winds are 
fundamental to the understanding of feedback processes of galaxies. 
In previous studies, the energy and metals of outflows are generally 
inferred from fitting observed spectra with thermal models.

The X-ray line emission, however, can arise not only from hot plasma,
but also from its interaction with
neutral cool gas. For example, 
highly ionised ions in the solar wind can readily pick up electrons from 
neutral species of a comet. 
These electrons, captured in excited states of the ions, will cascade down,
leading to X-ray line emission.
This process is called charge exchange X-ray emission 
(CXE, also called charge transfer) 
and explains the bright cometary X-ray emission \citep[e.g.,][]{Lis96, Cra97,
	Cra02}.
The process can be represented by:
\begin{equation}
A^{q+} + N \rightarrow A^{(q-1)+*} + N^{+},
\end{equation}
where a highly ionised ion $A^{q+}$ (like \Oe) picks up an electron 
from a neutral species $N$ (like H) and produces an excited ion
 $A^{(q-1)+*}$.
For historical studies and recent developments of the CXE, we refer
the readers to \citet{Den10} and references therein.

The charge-exchange cross-sections are quite large compared to the 
cross-sections of electron collisional excitation. For example, 
at collision velocities of a few hundred km/s, the charge-exchange
cross-sections of \Oe\ and \On\ are $\sim10^{-15}-10^{-14}$ cm$^2$, 
compared to those of electron collisional excitation 
($\sim10^{-20}-10^{-19}$cm$^2$). This makes the CXE important even 
though it takes place in a very thin region compared 
to the tenuous hot gas.

Different from the thermal emission, the CXE contributes only emission lines.
If the X-ray line emission, or part of it, is due to the CXE, the
measurement of the thermal and chemical properties of the hot outflow
based on thermal-only models will be misleading.
To correctly understand the hot outflow, it is crucial to determine the 
origin of the X-ray line emission 
and to separate the contributions from the CXE and the thermal emission.

With the unprecedented spatial resolution of \cha, the extra-planar diffuse
X-ray emission of a number of edge-on disk galaxies has been found to be correlated 
with H$\alpha$ emission \citep[e.g.,][]{Wang01, Cec02, Str00,Str04,Li08}; 
in less-inclined galaxies, 
X-ray emission is also found to be correlated morphologically with H$\alpha$ 
emission \citep[e.g.,][]{Tyl04}.
\citet{Lal04}
speculated on the importance of the CXE between the hot outflow and cool halo
gas for starburst galaxies. In the scenario of the CXE, H$\alpha$ emission
represents 
entrained cool gas clouds/filaments, while the X-ray emission arises from their 
interfaces with the surrounding hot plasma.
The spatial correlation between the H$\alpha$ and X-ray emissions is thus
naturally explained.

A marginal detection of an emission line at 0.459 keV is reported for the
X-ray cap above the disk of the starburst galaxy of M82, which may be due to 
the CXE of \CVI\ ($n=4\rightarrow1$) \citep{Tsu07}.
The direct evidence for the CXE occurring in M82 is 
from the line ratios of the \OVII\ K$\alpha$ triplet \citep{Ran08, Liu11}. 

The K$\alpha$ triplet of He-like ions is
a powerful diagnostic that can be used to reveal the origin of the X-ray line emission
\citep[for a recent review, see][]{Por10}.
It consists of a resonance line, two inter-combination lines, and a forbidden 
line. For a thermal plasma in ionization equilibrium, the electron collisional
excitation is efficient and 
favours the resonance line, while for the CXE, the electron downward cascade  
favours the triplet states and thus the forbidden line. 
Therefore, the line ratios of the triplet can be used to determine the origin of the X-ray
line emission. 

In a previous paper \citep{Liu11}, we  analysed the K$\alpha$ triplets of He-like ions
of \OVII, \NeIX, and \MgXI\ of M82 observed by the \xmm\ Reflection Grating Spectrometers
(RGSs).
The \OVII\ triplet is found to be dominated by the forbidden line, and the intensity ratio of the
\OVII\ triplet is consistent with an origin of the CXE.
The flux contributions of the CXE
are 90\%, 50\%, and 30\% to the \OVII, \NeIX, and \MgXI\ triplets, respectively.
The \OVII\ triplet was also used to show the possible presence of
charge exchanges between the hot gas and the cool gas spiral in the inner bulge of M31
\citep{Liu10}.

In this paper, we study the \OVII\ triplet for a sample of additional eight
star-forming galaxies, in comparison with M82. In all these galaxies, 
the forbidden lines are comparable to or stronger
than the resonance lines, which is in contrast to the thermal prediction.
We discuss the other possible mechanisms that can produce a relatively 
strong forbidden line, such as a collisional non-equilibrium-ionization
recombining/ionizing 
plasma and resonance scattering, which are found to be not favoured for our sample
galaxies.
It shows that the CXE may be a common process in galaxies with massive star formation.

The paper is structured as follows. We describe the observation sample 
in \S 2. The analysis results are presented in
\S 3. The conclusion and discussion are given in \S 4.
 Errors are given at 1 $\sigma$ confidence level.

\section{Observation sample }

\begin{table}
\caption{List of the sample} 
\begin{tabular}{lcccl}
\hline
name& obs ID &  $t_{\rm e}^a$ (ks) & d (Mpc)$^b$ &note$^c$  \\
\hline
M82 (NGC 3034) & 0206080101 & 62 & 3.9 & SB \\
& 		0560181301 &25 & & \\
NGC 253 & 	0152020101 & 72 &3.2 & SB \\
& 		0125960101 &43 & & \\
M51 (NGC 5194) & 	0303420101 & 38 &8.0 & spiral/nuSB\\
& 		0303420201 &27 & & \\
M94 (NGC 4736)& 0404980101 & 42 & 5.0 & spiral/nuSB\\
& 		0094360601 &14 & & \\
M83 (NGC 5236) & 0110910201 & 25 & 4.7 &bar-spiral/nuSB\\
NGC 2903 & 	0556280301 & 73 & 9.4 & bar-spiral/nuSB \\
M61 (NGC 4303) & 0205360101 & 26 & 12.1 & bar-spiral/nuSB\\
NGC 4631 & 0110900201 & 42 & 6.7 &spiral/nuSB\\
Antennae & 	0085220201 & 38 & 21.6& SB \\
& 		0500070401 &17 & & \\
\hline
\end{tabular}
\begin{description}
\item[] 
Note: $^a$ $t_{\rm e}$ is the useful exposure time after removing intense flare periods.
$^b$ The distance is the mean distance adopted from NED (NASA/IPAC extragalactic
database). 
$^c$ SB denotes for starburst and nuSB denotes for massive nuclear star formation.
M82, NGC 253, and NGC 4631 are edge-on galaxies, while others are close to face-on.
\end{description}
\end{table}

\begin{figure*}
\includegraphics[height=4.0in]{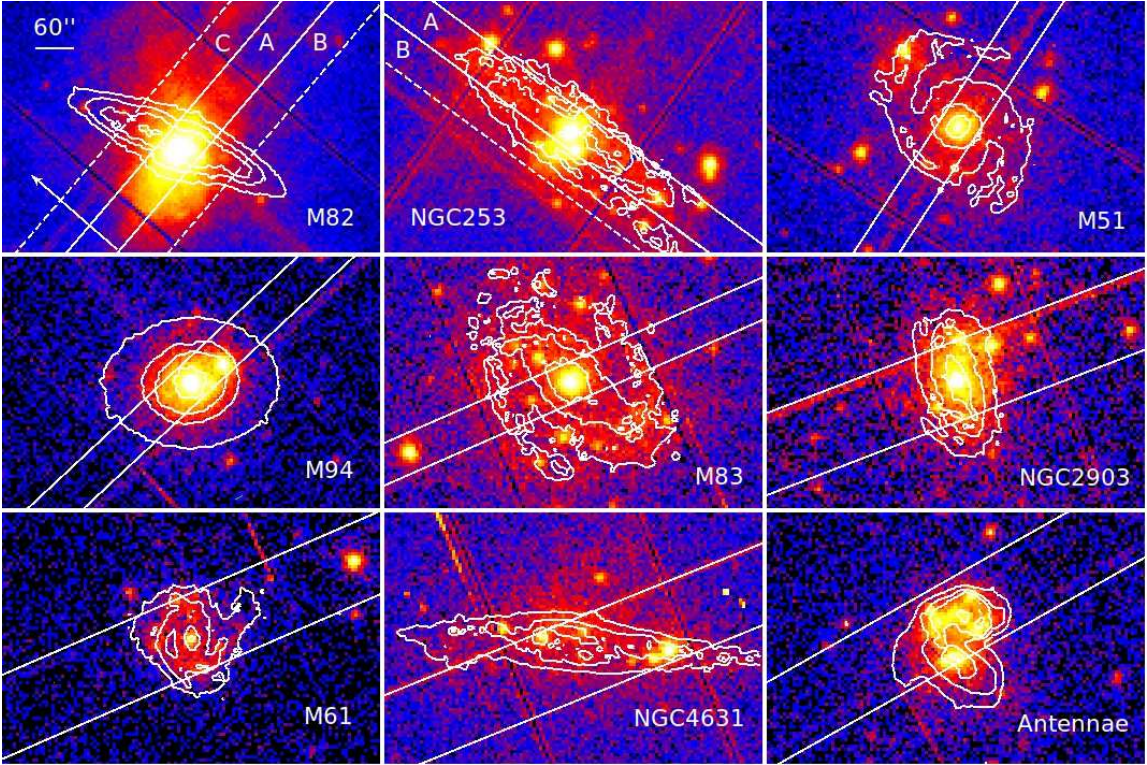}
\caption{ Soft X-ray images (0.5-2 keV) of the sample galaxies from \xmm\ EPIC-pn
data. For galaxies with multiple exposures, only the longest exposure is used.
Intense flare periods are removed and no background subtraction is performed.
The contours show the corresponding morphology for DSS-2 red images. 
The central RGS extraction regions (denoted as A for M82 and NGC 253)
are enclosed by the two solid lines in each panel.
The dashed lines and their adjacent solid lines of M82 and NGC 253
enclose the off-disk spectral regions, which are denoted as B and C. 
The arrow in the panel of M82 indicates the positive cross-dispersion 
direction.
}
\end{figure*}

The \xmm\ RGSs \citep{den01} are unique to provide high-resolution spectra for extended sources.
Besides M82, we select eight star-forming galaxies from the \xmm\ RGS data archive, 
which are bright enough for the analysis of the \OVII\ triplet. We exclude
the ones identified as Seyfert galaxies, for which photoionization by AGN may
be important. Thus, we have in total a sample of 
nine galaxies---three starburst-dominated galaxies and six more normal spirals with 
nuclear starburst regions. They are listed in Table 1. 

There are two RGSs on-board \xmm. CCD4 of RGS2 covering
the \OVII\ triplet failed early in the mission. Thus only RGS1 data are used
in the present work.
For those having multiple exposures (M82, NGC 253, M51, M94, and the Antennae),
we use the two longest exposures with similar observational configurations.
The observation IDs are listed in Table 1.
The \xmm\ software Science Analysis System (SAS, version 10.0) is used for
the reduction of photon events.

For extended sources, emission lines are broadened according to the relation
$\Delta\lambda=0.14\theta$ \AA, where the angular extent of the source $\theta$
is in units of arcmin. To obtain sufficient spectral resolution, 
for bright galaxies (M82, NGC 253, M51, M94, and M83) we limit the spectral extraction
to events within 60$''$ distance of the cross-dispersion direction.
For faint galaxies (NGC 2903, M61, NGC 4631, and the Antennae),
the extraction regions are 120$''$ distance of the cross-dispersion direction.
The corresponding extraction regions are outlined on the 0.5-2 keV European
Photon Imaging Camera (EPIC) pn images of the galaxies in Fig. 1.

We find that M82 shows qualitatively different characteristics of the \OVII\ triplet
at different cross-dispersion directions, which was not noticed in previous
studies. This spatial variation of the \OVII\ triplet of M82 is discussed in \S 3.2. 
For the disk galaxy NGC 253, which has sufficient counting statistics in the 
off-disk region, we also extract the spectra of
the off-disk region (60$''$ in the cross-dispersion direction) 
represented by the dashed line (denoted as B) in Fig. 1.  
As NGC 253 is offset from the RGS center, only one off-disk side is 
useful (see Fig. 5).

\section{Analysis and Results}

\begin{figure*}
\includegraphics[height=3.8in]{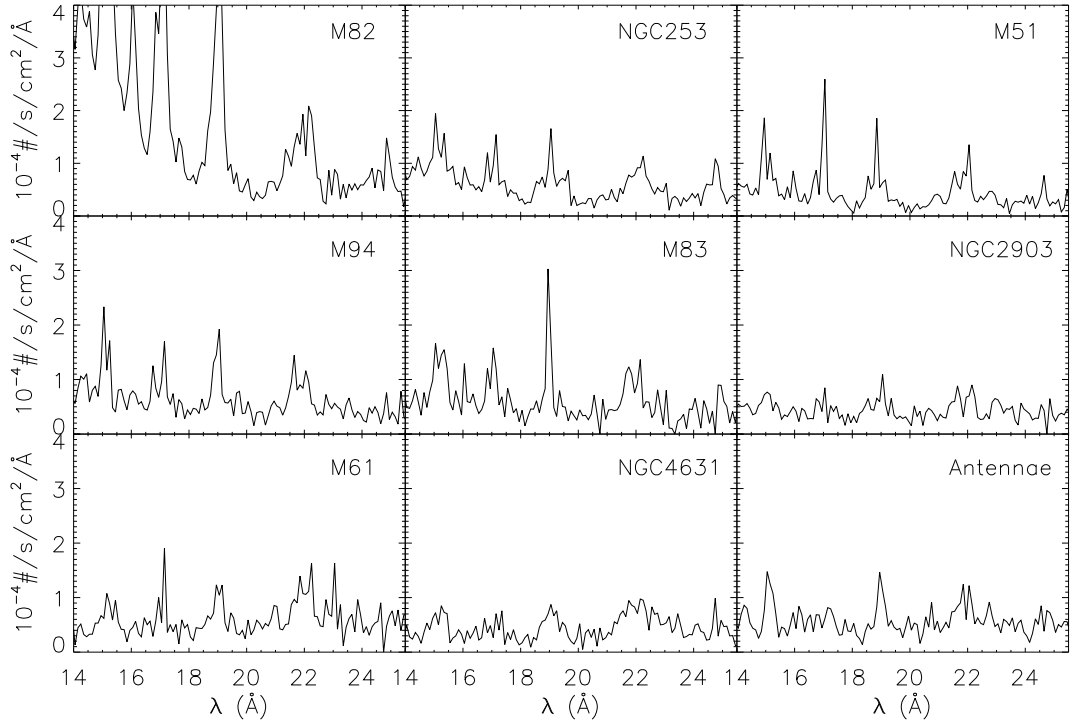}
\caption{\xmm\ RGS spectra of the central region of the sample galaxies. 
	Because CCD7 in the RGS1 covering
10-14 \AA\ failed, the wavelength range below 14 \AA\ is not shown. 
The spectra are dominated by the \OVIII\ \Lya line (19 \AA), the \OVII\ triplet
(22 \AA), and the \FeXVII\ lines (15 and 17 \AA).
}
\end{figure*}

The RGS spectra of the central region of the sample galaxies, corrected for the 
effective area, are plotted in Fig. 2. 
The emission lines are prominent in all galaxies and generally dominate the 
spectra. 
The emission lines are not 
necessarily from the thermal emission of the hot gas. Below we analyse their 
\OVII\ triplets, the intensity ratios of which are sensitive to the origin
of the X-ray line emission.

\subsection{Line ratios of the \OVII\ triplets}
\begin{figure*}
\includegraphics[height=5.5in]{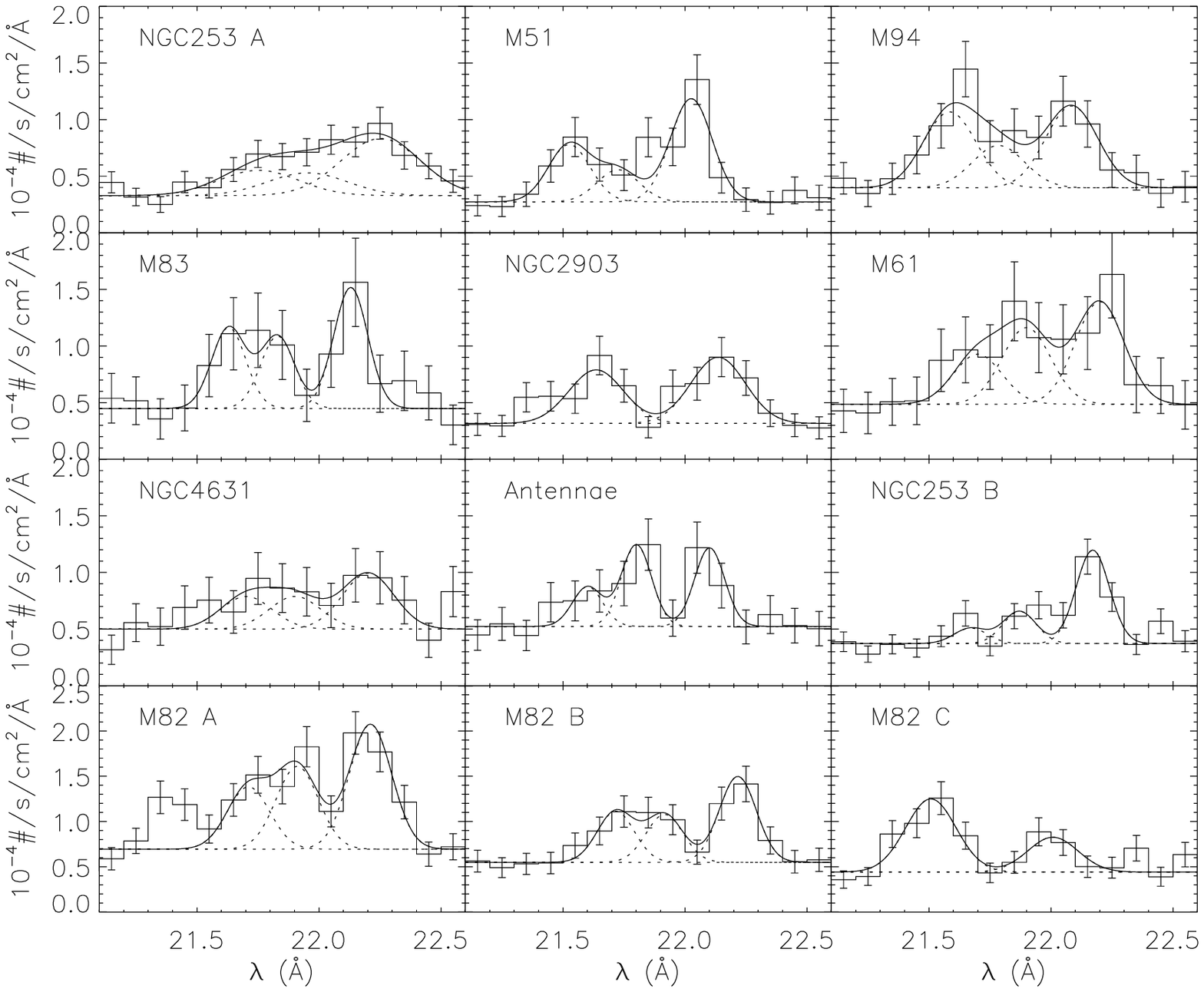}
\caption{\OVII\ triplets and the
best-fit model (solid lines) of three Gaussians (dotted lines)
	and a constant continuum.}
\end{figure*}

\begin{table*}
\caption{Results of fitting three Gaussians to the triplets} 
\begin{tabular}{lccccccccc}
\hline
name& $f_r^a$ & $f_i^a$ & $f_f^a$ & $f_{\rm con}^a$ & 
 $\Delta\lambda$(\AA)$^b$&
 $\sigma_{\lambda}$(\AA)$^b$&$\chi^2$/dof&$G=\frac{f+i}{r}$ &
	    ${\foiii/\foii}^c$ \\
\hline
NGC 253 A&0.9$\pm$0.6&0.8$\pm$0.4&2.1$\pm$0.6	& 3.3$\pm$0.5 &0.15$\pm$0.06
&0.16$\pm$0.06&5.8/9&3.2$\pm$1.5& 1.5$\pm$0.1\\
M51  	&1.1$\pm$0.5&0.6$\pm$0.8&2.0$\pm$0.6 	& 2.7$\pm$0.7 &
-0.07$\pm$0.02&0.09$\pm$0.02&8.2/9&2.4$\pm$1.4& 1.2$\pm$0.2\\
M94		&1.8$\pm$0.7&1.0$\pm$0.5&1.9$\pm$0.6 	& 4.0$\pm$0.9 &
-0.01$\pm$0.02&0.11$\pm$0.03&4.6/9&1.6$\pm$0.8& 1.0$\pm$0.1\\
M83 		&1.3$\pm$1.1&1.1$\pm$0.7&1.9$\pm$0.7 	& 4.5$\pm$1.0 &0.03$\pm$0.05
&0.07$\pm$0.04&3.0/9&2.3$\pm$1.5& 1.6$\pm$0.3\\
NGC 2903 &1.3$\pm$0.4&0.0$\pm$0.2&1.6$\pm$0.5   & 3.2$\pm$0.6 &0.03$\pm$0.03
&0.11$\pm$0.03&9.5/9&1.2$\pm$0.4& 1.0$\pm$0.1\\
M61      &1.1$\pm$1.2&1.7$\pm$0.8&2.3$\pm$1.0   & 4.9$\pm$1.3 &0.10$\pm$0.04
&0.10$\pm$0.03&3.6/9&3.6$\pm$2.6& 1.0$\pm$0.2\\
NGC 4631 &0.8$\pm$1.0&0.8$\pm$0.5&1.3$\pm$1.0   & 5.0$\pm$1.1 &0.10$\pm$0.05
&0.11$\pm$0.03&7.3/9&2.6$\pm$2.1& 1.1$\pm$0.2\\
Antennae    &0.5$\pm$1.0&1.1$\pm$0.6&1.1$\pm$0.7   & 5.2$\pm$0.8 &0.01$\pm$0.05
&0.06$\pm$0.04&5.2/9&4.4$\pm$2.9& 1.0$\pm$0.2\\
NGC 253 B&0.4$\pm$0.2&0.6$\pm$0.3&1.5$\pm$0.4	& 3.8$\pm$0.4 &0.07$\pm$0.01
&0.07$\pm$0.01&12.0/9&5.3$\pm$1.9& 1.9$\pm$0.2\\
M82 A$^d$		&1.5$\pm$0.5&2.0$\pm$0.4&3.0$\pm$0.6 	& 7.0$\pm$0.8 &0.11$\pm$0.02 
&0.08$\pm$0.01&6.9/7&3.3$\pm$1.0& 2.3$\pm$0.2\\
M82 B		&1.1$\pm$0.4&1.0$\pm$0.3&1.8$\pm$0.4 	& 5.5$\pm$0.5 &0.12$\pm$0.02 
&0.08$\pm$0.01&2.0/9&2.5$\pm$1.0& 2.2$\pm$0.2\\
M82 C		&2.1$\pm$0.7&0.0$\pm$0.4&1.0$\pm$0.6 	& 4.4$\pm$1.0 &-0.09$\pm$0.02 
&0.10$\pm$0.02&11.7/9&0.5$\pm$0.5& 2.4$\pm$0.3\\
\hline
\end{tabular}
\begin{description}
\item[] 
Note: For the meanings of the symbols, see eq (2). $^a$$f_{r,i,f}$ and $f_{\rm con}$ are
in units of $10^{-5}$photons/s/cm$^2$.
$^b$$\Delta\lambda$ and $\sigma_{\lambda}$ are the wavelength shift and
dispersion, respectively.
$^c$$\foiii$ and $\foii$ refer to the continuum-subtracted fluxes within
18.5 -- 19.5 \AA\ and 21.4 -- 22.4 \AA, respectively.
$^d$ The two data points between 21.3 and 21.5 \A are neglected when fitting
the triplet of M82 A (see discussion in \S 3.2).

\end{description}
\end{table*}

The observed \OVII\ triplets of the galaxies are plotted in Fig. 3. 
For the edge-on galaxies of NGC 253 and NGC 4631, the \OVII\ triplets are 
heavily blended. The forbidden lines are relatively well separated for 
others. Except for the C region of M82, Fig. 3 shows that for all galaxies the 
forbidden lines are comparable to or stronger than the resonance lines.

To study the line ratios of the \OVII\ triplets quantitatively, we fit a 
model of three Gaussians 
and a constant continuum to each of the observed triplets, which is written as 
\begin{equation}
f_{\rm model}=\frac{1}{\sqrt{2\pi}\sigma_{\lambda}}\sum_{j=\rm r,i,f}
f_j\exp\left[-\frac{(\lambda-\lambda_j-\Delta\lambda)^2}
{2\sigma_{\lambda}^2} \right] + f_{\rm con}, 
\end{equation}
where $\lambda_{\rm r,i,f}$ (21.6, 21.8 and 22.1 \AA) are the wavelengths of the resonance, 
inter-combination, and forbidden lines, respectively, $f_{\rm r,i,f}$ and 
$f_{\rm con}$ the corresponding fluxes,
$\sigma_{\lambda}$ the dispersion, and  $\Delta\lambda$ the wavelength
shift. In total, we have 15 bins of data, with 6 parameters. 
The errors are estimated using the bootstrap method.
The wavelength dispersions are largely determined by the spatial extent, while
the wavelength shifts are mainly determined by the spatial offset
between the peak X-ray line emission region and the adopted nominal center.
As the \xmm\ RGS is slit-less, we can not remove point sources from the RGS
spectra. Therefore, the continua of the RGS spectra are dominated by
point sources and instrument background. As a result, we choose to
fit the triplets with a local continuum within a fitting region 
extending beyond the triplets. The fitted continua are generally close to the 
fluxes at the wings of the \OVII\ triplets.
The fitting results are plotted in Fig. 3 and listed in Table 2.

The often-used line ratio 
\begin{equation}
G=\frac{f_{\rm f}+f_{\rm i}}{f_{\rm r}},
\end{equation}
which is a diagnostic of temperature of thermal plasmas,
is calculated and listed in Table 2. 
Because the inter-combination lines are poorly constrained due to blending,
the ratio $R=f_{\rm f}/f_{\rm i}$ is not included.

For a thermal plasma in ionization equilibrium, the ratio $G$ is 
smaller 
than 1.2 for temperatures higher than 0.1 keV.
All the fitted $G$ ratios are larger than 1.2, except for M82 C, which 
is discussed in the next section.
Therefore, the majority of the \OVII\ triplets of the sample galaxies can not come 
from the ionization-equilibrium hot gas directly, but may originate from the CXE.

For the CXE, the experiment measured ratio $G\sim2.2$ \citep{Bei03}.
For galaxies of M51, M83, and M82 B, the triplets are relatively well resolved, and
the fitted $G$ ratios $\sim2.3-2.5$ are in good agreement with the predicted 
value of the CXE. For M94 and NGC 2903, the triplets are also relatively resolved, but 
their $G$ ratios are $\sim1.2-1.6$, which indicate multiple origins.
Except for M82 C and NGC 253 B, the $G$ ratios of other galaxies are consistent with the 
value of the CXE within 1 $\sigma$ errors.

\subsection{\OVII\ triplets of M82}

\begin{figure}
\includegraphics[height=2.in]{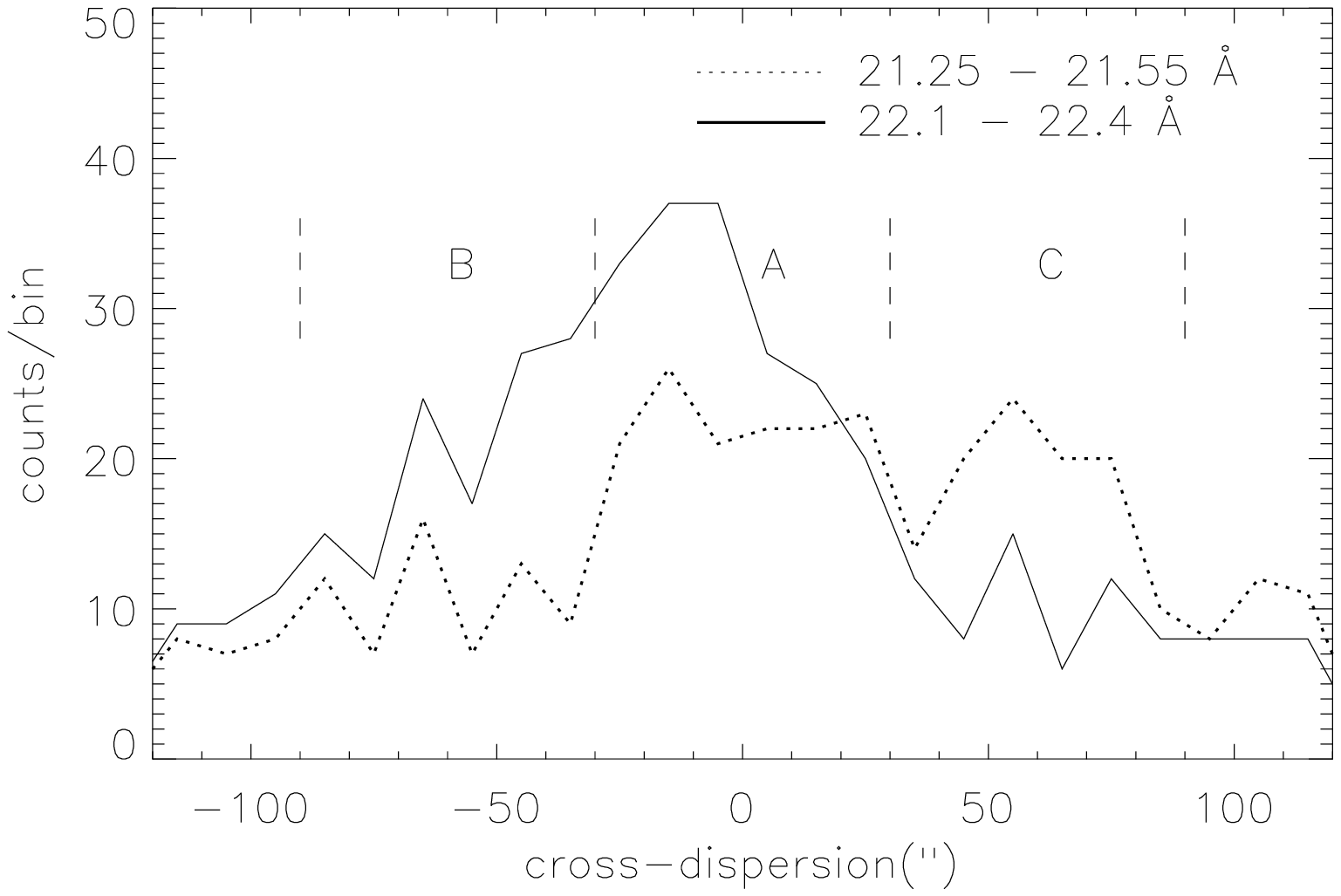}
\includegraphics[height=2.in]{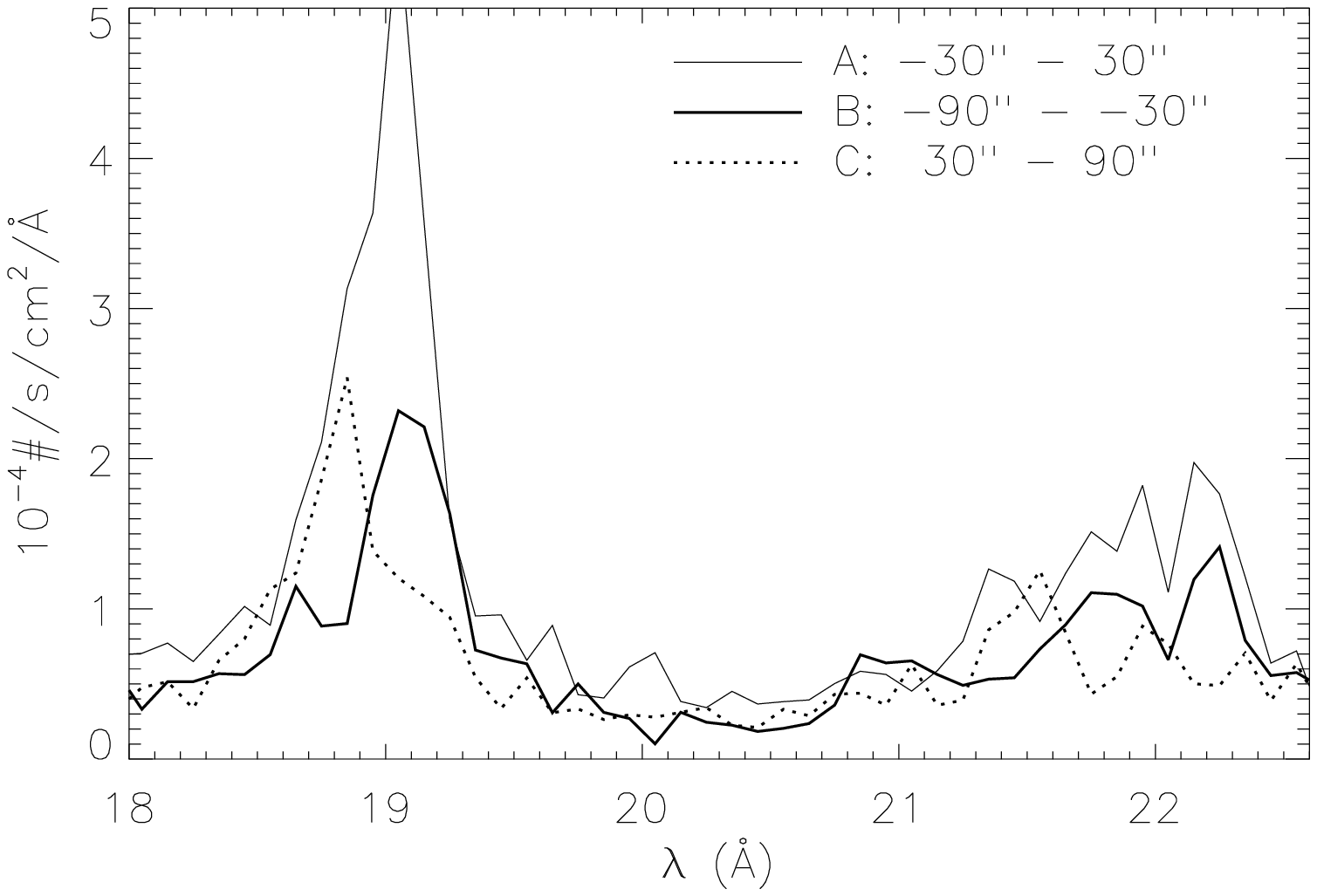}
\caption{Top: Comparison between the cross-dispersion distributions of M82 for 
photons between 21.25
and 21.55 \A and between 22.1 and 22.4 \AA, and marks of A, B, and C denote the three
regions adopted; Bottom: Comparison of the OVIII
\Lya lines and the \OVII\ triplets of M82
in three different regions. For clarity, errors are not plotted.}
\end{figure}

The observed triplet of M82 A in Fig. 3 shows a feature around 21.4 \AA, which can not
be fitted with the model of three Gaussians.
As shown in Fig. 1, for M82, the RGS cross-dispersion direction is not 
parallel to the outflow direction, and the \OVII\ triplet will have a different 
wavelength shift for different cross-dispersion position. 
Therefore, in the top panel of Fig. 4, we plot
the cross-dispersion distributions of M82 for photons between 21.25 and 21.55 \A\
and between 22.1 and 22.4 \AA. The photons between 22.1 and 22.4 \A\ are
peaked around the center, while the photons between 21.25 and 21.55 \A have a 
plateau from -30$''$ to 90$''$. Thus, we divide the cross-dispersion direction
into three regions: B (-90$''$ -- -30$''$), A (-30$''$ -- 30$''$), and C 
(30$''$ -- 90$''$), which are denoted in Fig. 1.
The \OVII\ triplets of the three regions are plotted in the bottom panel of
Fig. 4.

The \OVII\ triplets of the three regions, however, are not simply shifted according
to the spatial offsets, but have different line intensity ratios.
The peak around 21.5 \A of the C region can not be identified as the forbidden 
line of the \OVII\ triplets, otherwise, it will have a shift of 0.6 \AA,
which is larger than the shift provided by the spatial offset. Instead,
it is identified as the resonance 
line, as evidenced by the relative shift from the \OVIII\ line at 18.8 \AA.
For the B region, the forbidden line is dominant; for the C region, 
however, the resonance line is dominant.
The three lines peak at 21.4, 21.9, and 22.2 \A of the A region 
can not be fitted with a single triplet lines, but are a mixture of
features of B and C regions. As a result, we have neglected the two data
points between 21.3 and 21.5 \A when fitting the triplet of M82 A.
The different ratios of the \OVII\ triplets of different regions indicate 
complicated processes and multiple origins of the X-ray emission of M82.

%In our previous paper \citep{Liu11}, we used four sets of exposures,
%the dispersion directions of which are different from each other. 
%It diluted the distinction of the region C and we did not notice the variation.
%We check the H$\alpha$ emission of M82 in the corresponding B and C regions.
%H$\alpha$ emission is also enhanced in the B region compared to the C region,
%indicating abundant cool gas in the B region. This is 
%consistent with the explanation of the CXE origin of the relatively strong 
%forbidden line of B region. A detailed analysis of the spatial correlation of 
%H$\alpha$ and X-ray emission of M82 will be presented in a future work.

\subsection{Cross-dispersion profiles of the \OVII\ triplets}

\begin{figure*}
\includegraphics[height=3.in]{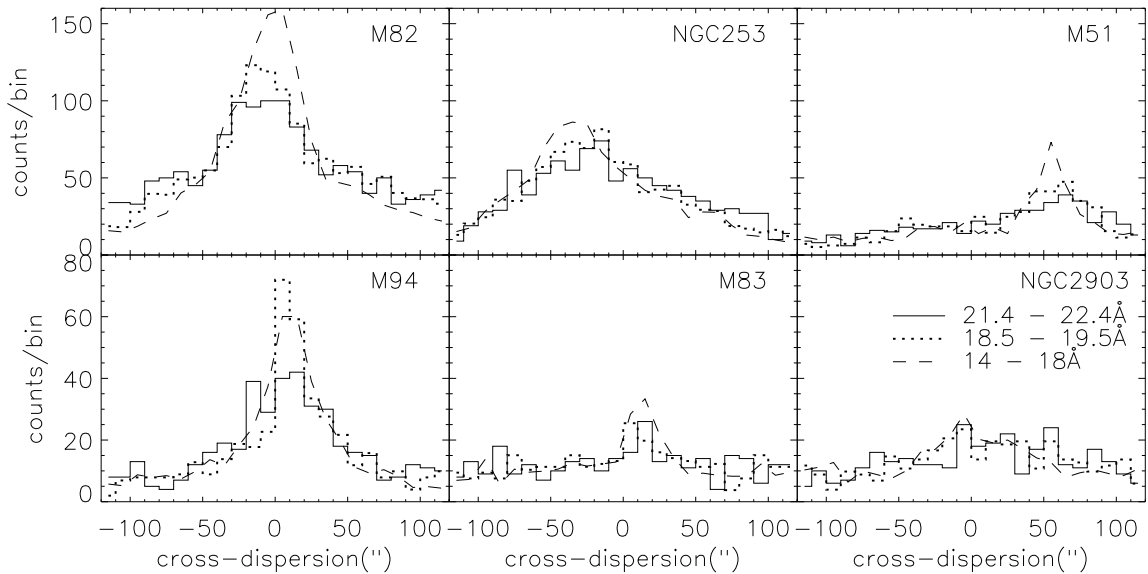}
\caption{Cross-dispersion profiles of the \OVII\ triplet (solid histogram), the 
\OVIII\ \Lya line (dotted histogram), and the emission between 14 and 18 \AA\ (dashed line). 
Both profiles of the \OVIII\ \Lya line and 
the emission between 14 and 18 \AA\ are normalized to the total counts of the \OVII\ triplet.
The \OVII\ triplets are for photons between 21.4 and 22.4 \AA\ and the 
\OVIII\ \Lya lines are for photons between 18.5 and 19.5 \AA.
}
\end{figure*}

While the origin of \OVII\ triplets can be tested through
their line ratios, we can not directly tell the origin of other lines 
such as the \OVIII\ \Lya line. However, we can still get some
hints by comparing the spatial distribution
of such lines with that of the \OVII\ triplet
in the cross-dispersion direction.

In Fig. 5, we plot the cross-dispersion distributions of 
the \OVII\ triplet (21.4--22.4 \AA), the \OVIII\ \Lya line (18.5--19.5 \AA), 
and the emission between 14 and 18 \A for six relatively bright galaxies.
We see that the profiles of the \OVIII\ \Lya line are similar to those of 
the \OVII\ triplet, except for M94, the \OVIII\ \Lya line of which is more centered.
The similarity between the distributions of the \OVIII\ \Lya line and the \OVII\ triplet 
indicates that they may have a similar origin.

The distributions of the emission between 14 and 18 \AA\ of the bright galaxies 
(M82, NGC 253, M51, and M94) are more centered compared to the corresponding 
profiles of the \OVII\ triplet. It may be due to the concentration of the
point sources of power-law continuum, as suggested by the increasing continua
at 14 -- 18 \AA\ in the spectra of Fig. 2.
Nevertheless, for NGC 253 and M51, the profiles of the emission between
14 and 18 \AA\ and the \OVII\ triplet appear to be similar outside the center. 

If most of the \OVIII\ \Lya line emission is indeed primarily from the CXE, it is
then interesting to
calculate the observed flux ratios of the \OVIII\ \Lya line to the \OVII\ triplet
($\foiii/\foii$), which
are listed in Table 2. The charge exchange cross sections 
are similar for \On\ and \Oe ions, and the efficiencies of producing a 
$n=2\rightarrow1$ photon
are also similar for the resulting excited \Oe\ and \Ov\ ions \citep{Gre01}.
If only a single charge exchange collision occurs (collisionally
thin case), the flux ratio $\foiii/\foii$ reflects the composition ratio
of \On/\Oe\ of the interacting hot gas. The observed flux ratios of
$\foiii/\foii$ are $\sim1-2$, which means the ratios of \On/\Oe\ are also 
$\sim1-2$ for the interacting hot gas. 
These ratios correspond to a temperature of $2.8-3.2\times10^6$ K in a
thermal equilibrium plasma.

\section{Conclusion and discussion}

We have analysed the \OVII\ triplets of a sample of nine nearby star-forming galaxies.
All the observed $G$ ratios (eq.3)
of the galaxies are larger than 1.2, which is the upper limit for a thermal plasma
with temperatures higher than 0.1 keV. For NGC 253, M51, M83, M61, NGC 4631,
and the Antennae, the observed
$G$ ratios are consistent with the laboratory measured ratio of the
CXE, while for M94 and NGC 2903, the observed $G$ ratios indicate multiple
origins. For M82, the \OVII\ triplets of different regions show different
line ratios. The cross-dispersion profiles
of the \OVIII\ \Lya line are similar to those of the \OVII\ triplet, 
indicating that the \OVIII\ \Lya lines may have a similar origin.
Below we discuss the CXE and other possible mechanisms for the \OVII\ triplet emission
of a high $G$ ratio.

\subsection{AGN contribution}

In our sample, three galaxies (M51, M61, and M94) show low level AGN
activities, which can complicate the interpretation of the 
\OVII\ triplets. A high $G$ ratio is also expected from the recombining ions due to
the photoionization by an AGN.
However, at least for M61 and M94, the analysis of the \cha\ images indicates
that the AGN is weak and unlikely to be important.
The \cha\ image of the nucleus of M61 reveals that the soft X-ray emission
traces an ultraviolet star-forming spiral down to an unresolved core
(1.5$''$), and the soft X-ray emission of the core accounts for only 15\% of the
total soft X-ray emission in the central 8$''$ region \citep{Jim03}.
For M94, the X-ray emission is resolved into
a diffuse component and a cluster of stellar X-ray sources, which appear
to follow a recent episode of starburst, and shows no evidence of an AGN
\citep{Era02}.

\subsection{Non-equilibrium-ionization model}

A collisionl non-equilibrium-ionization (NEI) recombining/ionizing plasma
can also produce a relatively strong forbidden line \citep[e.g.,][]{OP04}.
In a NEI recombining plasma, the recombination lags the cooling
\citep[e.g.,][]{BS99}, and the delayed recombination can produce a 
strong forbidden line. For the star-forming galaxies studied here, 
however, the X-ray emission is spatially correlated with the filamentary 
H$\alpha$ emission. This fact is consistent with the scenario that 
X-ray emission arises from the interfaces between the hot outflow and the
entrained cold gas, 
but would be difficult to explain for the NEI model in which the emission 
should come from the bulk of the outflow, 
where the low density regions will be the most over-ionized.		 

Furthermore, most of the \OVII\ triplet emission arises from the central
region of the galaxies.
According to \citet{CC85}, $T\propto r^{-4/3}$ in an adiabatic free-flow. 
The initial temperature is generally estimated to be around $10^8$ K
\citep[e.g.,][]{Vei05}. 
If we adopt a star-forming region of 100 pc (as in the case of M82), 
when it extends to 0.5 kpc, the temperature decreases a factor of 10, still
far above the fully stripping temperature of \On\ ion ($4\times10^6$ K).
A mass-loading factor of 10 will help the plasma to cool to 
about $10^6$ K, the peak \Ov\ ion temperature. But the mass-loading
leads to a high density, hence an equilibrium condition.
Thus, especially in the central regions where most of the \OVII\ triplet
emission arises from, a NEI recombining plasma seems unlikely to be important.

It is also possible to produce a high $G$ ratio in an ionizing NEI plasma
\citep[e.g.,][]{L99}, such as supernovae remnants (SNR), in which the heating 
generally precedes the ionization.
The inner-shell collisional ionization of \Os\ ions can lead to excited
\Ov\ ions and to the enhanced forbidden line emission.
However, this happens only in a short transition region when the \Os\ fraction is
large. In fact, the observed \OVII\ triplets of SNR generally show a thermal-like 
ratio \citep[e.g.,][]{Ras01,Beh01}. For the observed (accumulated plasma) spectra
of the sample galaxies, the \OVIII\ \Lya emission is stronger than the 
\OVII\ triplet, which means that the \Os\ ion fraction is negligible. 

\subsection{Resonance scattering}

Another process that may have effect on the line ratios is resonance
scattering. The resonance line has a larger optical depth due to its large
oscillator strength compared to the forbidden line. If the optical depth is
large enough,
the resonance line intensity can then be re-distributed from the central 
optical thick region to 
the outer optical thin region \citep{Gil87}, and the ratio $G$ is increased in the 
central 
region and reduced outside. For the off-disk \OVII\ triplets of M82 B and NGC 253 B, 
the $G$ ratios are higher than the ratio of thermal models, which can
not be due to resonance scattering. 
Furthermore if resonance scattering is important for the \OVII\ 
resonance line, so is for the \OVIII\ \Lya line. For thermal models, 
the ratio of $\foiii/\foii$ of $1-2$ corresponds to an ion ratio of
\Oe/\Ov $\sim4-7$,
which means the scattering optical depth of the \OVIII\ \Lya line is $1.6-3$
times larger than that of the \OVII\ resonance line. Thus, 
if resonance scattering is important, the profiles of the \OVIII\ \Lya
line should be flatter than those of the \OVII\ triplet,
which is not observed in Fig. 5.

\subsection{CXE model}

For the CXE, we can estimate the required property of the hot gas 
assuming that all the oxygen ions are in \Oe\ state and can participate in the
exchange.
Then, the \OVII\ triplet flux is $4\pi R^2Vn_{O^{7+}}/(4\pi D^2)$, where
$R$ is the emission radius and assumed to be 0.5 kpc, $V$ the velocity (1000
km s$^{-1}$), $D$ the distance (4 Mpc for M82), and $n_{O^{7+}}$ the \Oe\ ion density.
Taking the solar oxygen abundance ($5\times10^{-4}$), we find that an electron density of
$n_e\sim0.1$ cm$^{-3}$ is enough to produce the observed central \OVII\ triplet flux of 
M82 ($7\times10^{-5}$ photons cm$^{-2}$s$^{-1}$).
The effective interacting surface area can, in principle, be
substantially greater than $4\pi R^2$, depending on the topology of the cool gas
and the dynamics of the turbulent mixing of the hot and cool gases. 
The lack of spatial information limits further estimates, but
the details of the CXE process can be better constrained 
with future spatially-resolved high-resolution X-ray spectroscopy.

%The \OVII\ triplets of M82 show interesting spatial variations along the 
%cross-dispersion direction. One possible explanation is that 
%the outflow in the C region of M82 has a lower temperature of $\sim2\times10^6$ K,
%which is only efficient for the thermal emission, not the
%CXE. If the central star-forming region is offset the disk center towards 
%the negative direction of the cross-dispersion, then the positive outflow 
%will have a larger mass-loading factor and a lower temperature. 
%Further \xmm\ RGS observation with the dispersion direction perpendicular to
%the minor axis will test this scenario. Future spatially-resolved
%X-ray spectroscopy will also help to reveal the nature of the spatial 
%variation of the \OVII\ triplets of M82.

Since the majority of the \OVII\ triplets are consistent with the CXE and 
the emission between 14 and 20 \AA\ is likely to have a similar origin, 
one can not measure the
temperature and metal abundances of the hot gas assuming thermal models alone.
Instead, one needs to incorporate the CXE in the spectral modelling. 
The relative abundances of different ions can be obtained by comparing the
observed spectra with the CXE models, as has been done for 
the solar wind based on X-ray spectra of comets \citep[e.g.,][]{SC00,Kha03}.
For M82, the relative abundances
of O, Ne, and Mg were found to be similar to the solar values based on the CXE 
fluxes of their K$\alpha$ triplets \citep{Liu11}.
In some galaxies, one may need to consider the CXE
and the thermal emission simultaneously, as shown 
by the $G$ ratios of M94 and NGC 2903 and the \OVII\ triplets of M82. 
For such a task, a detailed modelling of the CXE is needed.
The CXE cross sections 
of heavy elements of Ne, Mg, and Fe need to be calculated and/or experimentally
measured.

\section*{Acknowledgements}

We appreciate our referee's critical and constructive report.
This research has made use of \xmm\ archival data.
XMM-Newton is an ESA science mission with instruments
and contributions directly funded by ESA Member States and the USA (NASA).
We also use images from DSS-2 (POSS-II), which was made by the California 
Institute of Technology with funds from the National Science Foundation,
the National Geographic Society, the Sloan Foundation,
the Samuel Oschin Foundation, and the Eastman Kodak Corporation. 
JRL and SM thank the Chinese Academy of Sciences for financial support and
QDW acknowledges the support by NASA via the grant NNX10AE85G.

\bibliographystyle{mn2e}

\end{document}